\documentclass[lettersize,journal]{IEEEtran}
\usepackage{amsmath,amsfonts}
\usepackage{algorithmic}
\usepackage{algorithm}
\usepackage{array}
\usepackage[caption=false,font=normalsize,labelfont=sf,textfont=sf]{subfig}
\usepackage{textcomp}
\usepackage{stfloats}
\usepackage{url}
\usepackage{verbatim}
\usepackage{graphicx}
\usepackage{cite}
\newtheorem{remark}{Remark}
\usepackage{amssymb}
\usepackage{booktabs}
\usepackage{xcolor}
\newcommand{\blue}[1]{{\textcolor[rgb]{0,0,1}{#1}}}

\begin{document}

\title{Iterative Semantic Decoding for Short Block Codes}

\author{Jiafu Hao, \IEEEmembership{Student Member, IEEE},
        Chentao Yue, \IEEEmembership{Senior Member, IEEE},
        Mingcheng Nie, \IEEEmembership{Student Member, IEEE},\\
        Branka Vucetic, \IEEEmembership{Fellow, IEEE},
        and Yonghui Li, \IEEEmembership{Fellow, IEEE}%
\thanks{Jiafu Hao, Chentao Yue, Mingcheng Nie, Branka Vucetic and Yonghui Li are with the School of Electrical and Computer Engineering, The University of Sydney, Sydney, NSW 2006, Australia
(e-mail: \{jiafu.hao; chentao.yue; mingcheng.nie; branka.vucetic; yonghui.li\}@sydney.edu.au). (\textit{Corresponding author: Chentao Yue})}%

\thanks{Code available: \protect\url{https://github.com/Jeh100/Iterative-SEC}. The work of Chentao Yue was supported by ARC under Grant DE250101332.}%
}

\maketitle

\begin{abstract}
This paper proposes an iteratively enhanced semantic receiver for natural-language text transmission over noisy wireless channels using multiple short block codes. At the transmitter, each sentence is permuted by a character-level interleaver, partitioned into segments, and independently encoded by short block codes. At the receiver, we develop an iterative decoder consisting of a channel decoder and a language model, where a de-interleaver between them disperses the burst decoding errors within each segment across the sentence. In each iteration, the language model denoises the channel decoding output, and the denoised characters verified to be consistent with the channel observations are fed back to the channel decoder as semantic information for the next iteration. Simulation results on the Stanford Natural Language Inference (SNLI) corpus over the additive white Gaussian noise (AWGN) channel show that the proposed receiver achieves approximately $1.5$~dB block error rate (BLER) gain over conventional short-block coding, while maintaining BLEU and ROUGE scores above $99\%$ at SNRs beyond $1.0$~dB.
\end{abstract}

\begin{IEEEkeywords}
Short block codes, semantic error correction, language model, iterative decoding, ordered statistics decoding.
\end{IEEEkeywords}

\section{Introduction}
\label{sec:intro}
\IEEEPARstart{S}{emantic} communication (SemCom) has emerged as one of the key paradigms for next-generation wireless networks, shifting the design objective from accurate bit-level recovery toward semantic completeness~\cite{survey}. The core idea of SemCom is to incorporate source-level knowledge, such as linguistic patterns\blue{,} into the communication process. Among the approaches developed under this paradigm, joint source--channel coding (JSCC) employs a pair of neural networks to jointly optimize source compression and channel error protection, demonstrating better robustness than conventional coding schemes in applications such as wireless image transmission \cite{cv_jscc} and natural-language transmission \cite{nlp_jscc}.

However, adopting JSCC requires substantial modifications to the conventional separate source and channel coding architecture~\cite{survey}, resulting in considerable deployment costs. Moreover, JSCC models are typically optimized for specific channel conditions, exhibiting poor generalization when the operating environment changes~\cite{10436878}. These limitations are particularly restrictive in systems with dynamic channel conditions and service requirements, where JSCC models would require frequent retraining. A representative example is the non-terrestrial network (NTN), in which low-Earth-orbit (LEO) satellites support short-packet transmissions under stringent power constraints and low signal-to-noise ratio (SNR) conditions \cite{3GPP,leo}; in this regime, short blocklengths inherently incur a reliability penalty \cite{PPV}. These constraints motivate a receiver-side augmentation of conventional short-block decoding to preserve compatibility with existing wireless standards while exploiting semantic information to compensate for channel impairments.

The semantic error correction (SEC) paradigm augments conventional systems with language-model-based semantic priors, allowing the receiver to exploit linguistic context either after channel decoding or directly within the decoding process. In particular, the segmented SEC scheme~\cite{sec} encodes a sentence with multiple short block codes, so that the correctly decoded segments provide linguistic context for a language model to recover the erroneous ones. Building on this architecture, semantic list decoding (SLD)~\cite{sld} generates multiple candidate reconstructions for each unreliable segment and selects the candidate most consistent with the received soft information. Other variants exploit semantic priors in most-reliable-basis selection~\cite{sem_osd}, trellis search~\cite{llm_viterbi}, and Bayesian reliability combining~\cite{cl_sec}. TokCom~\cite{token_sec} designs token-to-bit mappings at the transmitter to make channel errors less semantically destructive.
 However, all these methods apply semantic correction in a single pass and do not feed the semantic information back to the channel decoder for further refinement.

In this paper, we propose an iterative SEC framework that establishes a feedback loop between the language model and the channel decoder. At the transmitter, the sentence is permuted by a character-level interleaver, partitioned into segments, and independently encoded by short block codes. At the receiver, an iterative decoder alternates between ordered statistics decoding (OSD) and language-model denoising. In each iteration, the decoded segments are de-interleaved, which disperses the burst errors within any erroneous segment into isolated character errors across the sentence. The language model then denoises the resulting sentence to produce a semantic estimate. A character-level verification module retains only the estimated characters consistent with the corresponding channel log-likelihood ratios (LLRs). The retained characters are then pinned as known constraints in the next decoding iteration by adding a large signed bias to their LLRs. Simulation results on the Stanford Natural Language Inference (SNLI)~\cite{snli} corpus over the additive white Gaussian noise (AWGN) channel show that the proposed design outperforms SEC by approximately $1$~dB and the conventional short-code scheme by approximately $1.5$~dB in block error rate (BLER), reaching a BLER of $10^{-4}$ at $2.5$~dB. Moreover, it maintains bilingual evaluation understudy (BLEU)~\cite{bleu} and recall-oriented understudy for gisting evaluation (ROUGE)~\cite{rouge} scores above $99\%$ beyond $1.5$~dB.


\section{System Model}
\label{sec:system_model}

\subsection{Multiple Short Linear Block Codes}
\label{subsec:channel_coding}

We consider the transmission of a natural-language sentence $\mathbf{s}$ of character length $\ell$ over a noisy channel. Each character is first mapped to an 8-bit American Standard Code for Information Interchange (ASCII) representation, yielding a bitstream of total length $8\ell$. Rather than encoding the entire sentence as a single long codeword, the multiple short code (MSC) framework partitions $\mathbf{s}$ into $q$ equal-length segments of $\ell_{\mathrm{MSC}}=\ell/q$ characters. This process yields $q$ bitstreams $\{\mathbf{b}_{1},\ldots,\mathbf{b}_{q}\}$, each containing $k=8\ell/q$ bits, which are then independently encoded and transmitted.

Specifically, each segment $\mathbf{b}_{i}$ is encoded by a binary linear block code $\mathcal{C}_{\mathrm{MSC}}(n,k)$, which maps $k$ information bits into an $n$-bit codeword ($n>k$) through the generator matrix $\mathbf{G}_{\mathrm{MSC}}\in\{0,1\}^{k\times n}$, i.e., $\mathbf{c}_{i}=\mathbf{b}_{i}\mathbf{G}_{\mathrm{MSC}}\in\{0,1\}^{n}$. Note that we adopt the systematic form $\mathbf{G}_{\mathrm{MSC}}=[\mathbf{I}_k,\mathbf{P}]$, so that the first $k$ positions in a codeword directly carry the transmitted ASCII information bits and the remaining $n-k$ positions serve as parity bits for error protection. The codeword is then binary phase-shift keying (BPSK) modulated as $\mathbf{x}_{i}=1-2\mathbf{c}_{i}\in\{-1,+1\}^{n}$ and transmitted over an AWGN channel, yielding the received signal
$$\mathbf{y}_{i}=\mathbf{x}_{i}+\mathbf{z}_{i}.$$
Here, $\mathbf{z}_{i}\sim\mathcal{N}(\mathbf{0},\sigma^{2}\mathbf{I}_{n})$ is AWGN with variance $\sigma^{2}$ and the SNR is defined as $1/\sigma^{2}$.  At the receiver, all received vectors $\mathbf{y}_{i}$, $i=1,\ldots,q$, are subsequently converted into LLRs as $\boldsymbol{\lambda}_i=2\mathbf{y}_i/\sigma^{2}$, and decoded in parallel by a near-maximum-likelihood (ML) short-block decoder (e.g., OSD~\cite{OSD}), yielding the bit estimate $\hat{\mathbf{b}}_{i}$; a decoding error occurs whenever $\hat{\mathbf{b}}_{i}\neq\mathbf{b}_{i}$. The concatenated estimate $\hat{\mathbf{b}}=[\hat{\mathbf{b}}_{1},\ldots,\hat{\mathbf{b}}_{q}]$ is then mapped through ASCII source decoding to the sentence estimate $\hat{\mathbf{s}}=[\hat{\mathbf{s}}_{1},\ldots,\hat{\mathbf{s}}_{q}]$, with each $\hat{\mathbf{s}}_{i}$ of length $\ell_{\mathrm{MSC}}$. Since the $q$ segments are encoded and decoded independently, decoding failures do not propagate across segments, leaving the correctly decoded segments as reliable context for the semantic processing module in Section~\ref{subsec:semantic_module}.

\begin{remark}
The framework adopts fixed-length ASCII mapping rather than source compression, since compression is incompatible with the character-level interleaving in Section~\ref{sec:method}. Compressing before interleaving removes the character boundaries on which the interleaver operates, whereas interleaving before compression destroys the linguistic structure that compression exploits. The resulting loss is minor, as for short texts, compression methods such as DEFLATE adopted by 3GPP offer little rate saving over the fixed-length ASCII mapping.
\end{remark}

\subsection{Ordered Statistics Decoding}
OSD~\cite{OSD} is a near-ML soft-decision decoder for linear block codes. For BPSK transmission, each entry $\lambda_{i,j}$ of $\boldsymbol{\lambda}_i$ provides both the hard decision and its reliability: $h_j=0$ for $\lambda_{i,j}\geq0$ and $h_j=1$ otherwise, while $\alpha_j=|\lambda_{i,j}|$ measures the reliability of the decision. OSD exploits this information by sorting the received positions in descending order of $\alpha_j$ and applying the same permutation to the generator matrix. After Gaussian elimination, the generator matrix is transformed into systematic form $\tilde{\mathbf{G}}=[\mathbf{I}_k,\tilde{\mathbf{P}}]$, where the first $k$ positions form the most reliable basis (MRB). OSD then applies test error patterns (TEPs) to the hard decisions in the MRB, re-encodes the resulting hypotheses, and selects the candidate with the minimum weighted Hamming distance to the received soft information. Consequently, assigning a sufficiently large reliability magnitude to a bit position promotes it into the MRB and allows it to act as a highly reliable constraint in subsequent OSD decoding, which is the property exploited by the semantic feedback mechanism in Section~\ref{subsec:semantic_external_information}.


\subsection{Semantic Processing Module}
\label{subsec:semantic_module}

\subsubsection{Pre-trained Language Model}
The semantic module is built upon the Bidirectional and Auto-Regressive Transformer (BART)~\cite{bart}, which is a pre-trained sequence-to-sequence neural model that combines a bidirectional encoder with an autoregressive decoder. The BART encoder produces contextual representations of the entire input, on which the decoder autoregressively generates output tokens. BART is trained to reconstruct the original text from corrupted input sequences by minimizing a token-level cross-entropy loss, where the corruption is obtained through masking, deletion, and substitution. This denoising objective closely resembles the burst error correction task at the receiver, where channel-decoding failures appear as corrupted characters in the decoded text.

\subsubsection{SEC Pipeline}
The sentence estimate $\hat{\mathbf{s}}$ obtained after parallel channel decoding is fed into the SEC module~\cite{sec}, which operates as a sentence-level denoiser based on a fine-tuned BART model. Specifically, $\hat{\mathbf{s}}$ is first tokenized by the byte pair encoding (BPE) $f_{\mathrm{BPE}}(\cdot)$~\cite{BPE} as
\begin{equation}
    \hat{\mathbf{t}}=f_{\mathrm{BPE}}(\hat{\mathbf{s}})=[\hat{t}_{1},\hat{t}_{2},\ldots,\hat{t}_{\ell_{\mathrm{tok}}}],
    \label{eq:bpe}
\end{equation}
where $\ell_{\mathrm{tok}}$ is the token sequence length, which varies with sentence content. The token sequence $\hat{\mathbf{t}}$ is then passed to the BART-based SEC model, yielding the reconstructed sentence
\begin{equation}
    \tilde{\mathbf{s}}=f_{\mathrm{SEC}}(\hat{\mathbf{t}};\boldsymbol{\theta}_{\mathrm{SEC}}),
    \label{eq:sec_output}
\end{equation}
where $\boldsymbol{\theta}_{\mathrm{SEC}}$ denotes the fine-tuned model parameters. Note that $\tilde{\mathbf{s}}$ is an estimate of $\mathbf{s}$, whose length may differ from $\ell$ due to the autoregressive generation of BART; this length mismatch is handled in Section~\ref{sec:method}.

\begin{figure}[t]
   \centering
   \includegraphics[width=0.99\linewidth]{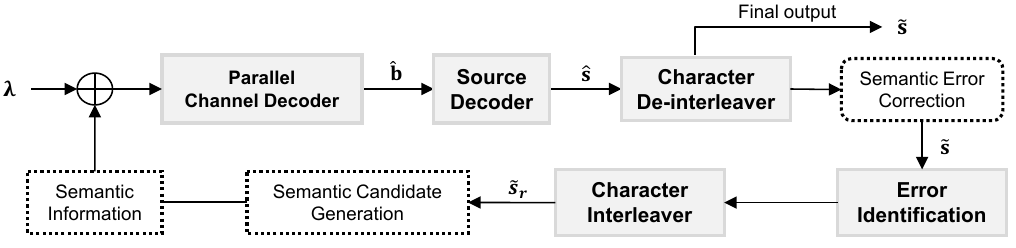}
   \caption{Proposed iterative receiver framework with parallel short block codes and SEC module. Modulation and demodulation are omitted for clarity.}
   \label{fig:system_model}
   \vspace{-0.8em}
\end{figure}

\section{Proposed Iterative SEC Framework}
\label{sec:method}
Building on the SEC scheme \cite{sec}, we propose an iterative framework, as depicted in Fig.~\ref{fig:system_model}. The transmitter interleaves the input sentence at the character level before forwarding it through the standard MSC pipeline, and the receiver iterates between OSD-based channel decoding and SEC-based semantic reconstruction. The design of each component is detailed below.


\subsection{Sentence Encoding with Character Interleaving}
\begin{figure}[t]
   \centering
   \includegraphics[width=0.99\linewidth]{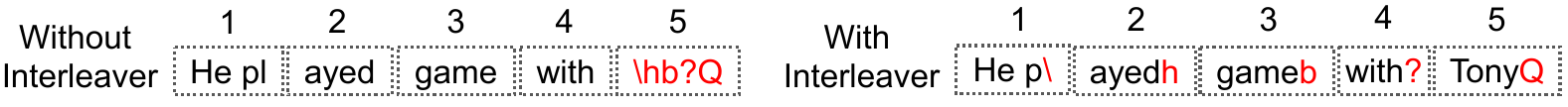}
    \caption{Effect of character-level interleaving on a segment decoding failure.}
   \label{fig:example}
   \vspace{-0.8em}
\end{figure}

At the transmitter, the sentence $\mathbf{s}$ is first permuted by a character-level interleaver $\Pi(\cdot)$, i.e., $\mathbf{s}'=\Pi(\mathbf{s})$, which arranges the $\ell$ characters into an $N_{1}\times N_{2}$ array with $N_{1}N_{2}=\ell$ and reads them out along the orthogonal dimension. Without interleaving, the decoding failure of segment $i$ corrupts a contiguous block of $\ell_{\mathrm{MSC}}$ characters in the reconstructed sentence $\hat{\mathbf{s}}$.
Fig.~\ref{fig:example} shows the sentence ``He played game with Tony.'' split into five segments, where corrupts the characters corresponding to the semantically critical word ''Tony''. With interleaving, the same errors are dispersed across the sentence after de-interleaving, so each affected word retains most of its characters, providing strong local evidence for SEC to combine with the surrounding context.
The permuted sentence $\mathbf{s}'$ is then processed through the MSC pipeline described in Section~\ref{subsec:channel_coding}, where all resulting quantities are marked with a prime, e.g., the transmitted segments $\mathbf{b}_{i}'$ and the LLRs $\boldsymbol{\lambda}_{i}'$.


\subsection{Semantic Feedback Candidate Generation}
At the receiver, each LLR vector $\boldsymbol{\lambda}_i'$ is decoded by OSD to obtain the bit estimate $\hat{\mathbf b}_i'$. Following the segment-level confidence evaluation in~\cite[Eq. (11)]{sld}, we evaluate for each segment the posterior confidence $P_{i}\in(0,1)$ that the OSD-decoded segment $\hat{\mathbf b}_i'$ is correct, computed from the correlation between the best OSD candidate codeword and the received vector $\mathbf{y}_{i}'$. Segments with low confidence are regarded as likely OSD decoding failures and are collected into the error pool
\begin{equation}
    \mathcal{S}_{\mathrm{err}} = \bigl\{i: P_{i}<T_{\mathrm{OSD}}\bigr\},
\end{equation}
where $T_{\mathrm{OSD}}\in(0,1)$ is a segment-level confidence threshold. If $\mathcal{S}_{\mathrm{err}}=\varnothing$, no segment is identified as unreliable, and no further SEC-based iterative processing is required.

The bitstreams $\hat{\mathbf b}_i'$ then undergo ASCII source decoding and de-interleaving via $\Pi^{-1}(\cdot)$, yielding the sentence estimate $\hat{\mathbf {s}}$. Subsequently, $\hat{\mathbf {s}}$ is passed to the
SEC module, which reconstructs the sentence:
\begin{equation}
    \tilde{\mathbf{s}}
    = f_{\mathrm{SEC}} \left( f_{\mathrm{BPE}}(\hat{\mathbf s}); \boldsymbol{\theta}_{\mathrm{SEC}} \right),
\end{equation}
where $\tilde{\mathbf{s}} = [\tilde{s}_{1}, \ldots, \tilde{s}_{r}, \ldots, \tilde{s}_{\ell_e}]$ is the reconstructed sentence with character length $\ell_e$.

The reconstructed characters in $\tilde{\mathbf{s}}$ corresponding to the segments in $\mathcal{S}_{\mathrm{err}}$ are selected as targets for semantic feedback in the next iteration. For these target characters, the verification module applies two rules in sequence and produces the feedback candidate set $\mathcal{V}$. The candidates in $\mathcal{V}$ are fed back to the next OSD iteration, as detailed in Section~\ref{subsec:semantic_external_information}.

\textbf{Rule 1: Substitution-only feedback.} BART produces $\tilde{\mathbf{s}}$ through token-by-token generation, and thus the output length is not tied to that of $\hat{\mathbf{s}}$. Aligning $\hat{\mathbf{s}}$ and $\tilde{\mathbf{s}}$ character by character yields four types of edits: substitution, insertion, deletion, and unchanged. Since the feedback must be injected into the OSD of the next iteration, only characters that are aligned with the transmitted bit positions and carry new information are useful. Insertions and deletions destroy the character alignment with the received sequence and thus cannot be injected. Unchanged characters are aligned but provide no new information. The following rule is therefore restricted to substitution positions.

\textbf{Rule 2: Channel-reliable candidate selection.} A substitution proposed by BART is trustworthy only if it does not contradict the channel observation. For each target character $\tilde{s}_{r}$, let $\pi(r)=(i,m)$ denote its transmitted position in the interleaved sequence, i.e., the $m$-th character of segment $i\in\mathcal{S}_{\mathrm{err}}$, and let $\boldsymbol{\lambda}_{i,m}'$ collect the LLRs at the corresponding systematic positions. Converting $\tilde{s}_{r}$ to its 8-bit ASCII representation $\tilde{\mathbf{b}}_{r}$ and the BPSK form $\tilde{\mathbf{x}}_{r}=1-2\tilde{\mathbf{b}}_{r}$, we define the per-character soft correlation score
\begin{equation}
    \rho_{r} = \tilde{\mathbf{x}}_{r}^{\top}\boldsymbol{\lambda}_{i,m}',
    \label{eq:soft_corr}
\end{equation}
which is large when $\tilde{s}_{r}$ agrees with the channel observation. The character is admitted if its score is sufficiently high, i.e.,
\begin{equation}
\mathcal{V} = \bigl\{(r,\tilde{s}_{r})\;:\;\rho_{r}\geq T_{\mathrm{h}}\bigr\},
\end{equation}
where $T_{\mathrm{h}}>0$ is a reliability threshold. If $\mathcal{V}=\varnothing$, the character with the largest $\rho_{r}$ is admitted only if $\rho_{r}\geq T_{\mathrm{low}}$, where $0<T_{\mathrm{low}}<T_{\mathrm{h}}$ is a secondary threshold, allowing a moderately reliable candidate to be fed back when no high-reliability candidate is available. If no candidate is admitted, the semantic-feedback iteration terminates and the current OSD output is retained.

\subsection{Semantic Information and Iterative Decoding}
\label{subsec:semantic_external_information}
The candidate set $\mathcal{V}^{(t)}$ produced at iteration $t$ is indexed in the sentence domain, whereas OSD operates on the interleaved segments. Each candidate $(r,\tilde{s}_{r})$ is therefore mapped back to the interleaver domain through $\pi(r)=(i,m)$ before injection. Let $\boldsymbol{\lambda}_{i}'^{(t)}$ denote the soft input of OSD at iteration $t$, initialized as $\boldsymbol{\lambda}_{i}'^{(0)}=\boldsymbol{\lambda}_{i}'$, and let $\boldsymbol{\lambda}_{i,m}'^{(t)}$ collect its entries at the systematic positions of the $m$-th character of segment $i$.

Each candidate $(r,\tilde{s}_{r})\in\mathcal{V}^{(t)}$ is injected as semantic hard information by updating
\begin{equation}
    \boldsymbol{\lambda}_{i,m}'^{(t+1)} = \boldsymbol{\lambda}_{i,m}'^{(t)} + M\,\tilde{\mathbf{x}}_{r},
    \label{eq:hard_injection}
\end{equation}
where $\tilde{\mathbf{x}}_{r}$ is the BPSK form defined in Rule~2 and $M$ is a sufficiently large value. These positions rank highest in the OSD reliability ordering and act as known constraints in the next decoding round. All other positions, including the parity bits, remain unchanged from iteration $t$.

The updated vectors $\{\boldsymbol{\lambda}_{i}'^{(t+1)}\}_{i=1}^{q}$ are decoded in parallel OSD to generate the new error pool $\mathcal{S}_{\mathrm{err}}^{(t+1)}$. The iteration terminates when $\mathcal{S}_{\mathrm{err}}^{(t+1)}=\varnothing$ or $t+1=I_{\max}$, where $I_{\max}$ is the maximum number of iterations, and the OSD-decoded sentence estimate is returned. Otherwise, the resulting bits are source-decoded, de-interleaved, and passed to the SEC module to generate $\tilde{\mathbf{s}}^{(t+1)}$.

\section{Simulation Results}
\subsection{Implementation Details}

\subsubsection{Dataset and Training} \label{section5train}
We use the SNLI corpus \cite{snli}, with about 19,000 sentences for training and 550 for testing. Sentence lengths $\ell$ range from 57 to 64 characters; each sentence is zero-padded to $64$ characters, i.e., $512$ bits after ASCII mapping, and the padding is removed before SEC processing. Besides, we set character-interleaving array with $N_1=q=8$ and $N_2=\ell/q=8$.

The training set $\mathcal{D}_{\mathrm{SEC}}=\{(\mathbf{s}^{(j)},\hat{\mathbf{s}}^{(j)})\}_{j=1}^{N}$ is constructed by transmitting each sentence over an AWGN channel multiple times at SNRs in $\{0, 0.5, 1, 1.5, 2, 2.5\}$~dB, followed by order-4 OSD decoding and segment-level confidence evaluation. Only sentences that contain at least one erroneously decoded segment are retained, yielding $N=100{,}000$ pairs for training. Based on the pretrained BART-base model, fine-tuning is performed by minimizing the token-level cross-entropy loss as
\begin{equation}
    \mathcal{L} = -\frac{1}{N}\sum_{j=1}^{N}\sum_{\tau=1}^{\ell_{\mathrm{tok}}^{(j)}}
    \log P\!\bigl(\mathbf{s}^{(j)}_{\tau}\bigm|\mathbf{s}^{(j)}_{<\tau},\hat{\mathbf{s}}^{(j)}\bigr),
\end{equation}
where $\mathbf{s}^{(j)}_{\tau}$ is the $\tau$-th target token of $f_{\mathrm{BPE}}(\mathbf{s}^{(j)})$ and $\ell_{\mathrm{tok}}^{(j)}$ is its token length. The AWGN channel and all coding pipelines are implemented in Sionna \cite{sionna}; fine-tuning runs on a single NVIDIA RTX 5090 GPU, with hyperparameters listed in Table~\ref{table:sim_params}.

\begin{table}[t]
\centering
\vspace{-1.1em}
\caption{Training and inference parameters.}
\label{table:sim_params}
\renewcommand{\arraystretch}{1.15}
\begin{tabular}{@{}l c | l c | l c@{}}
\toprule
\textbf{Param.} & \textbf{Value} & \textbf{Param.} & \textbf{Value} & \textbf{Param.} & \textbf{Value} \\
\midrule
Learning rate            & $3\times 10^{-5}$ & Batch size            & 128   & $I_{\max}$    & 5  \\
$M$        & 999               & $T_{\mathrm{low}}$\blue{,} $T_{\mathrm{h}}$      & 12.5\blue{,} 25             & $T_{\mathrm{OSD}}$   & 0.1\\
\bottomrule
\end{tabular}
\vspace{-1em}
\end{table}

\begin{figure*}[t]
   \centering
   \includegraphics[width=0.9\linewidth]{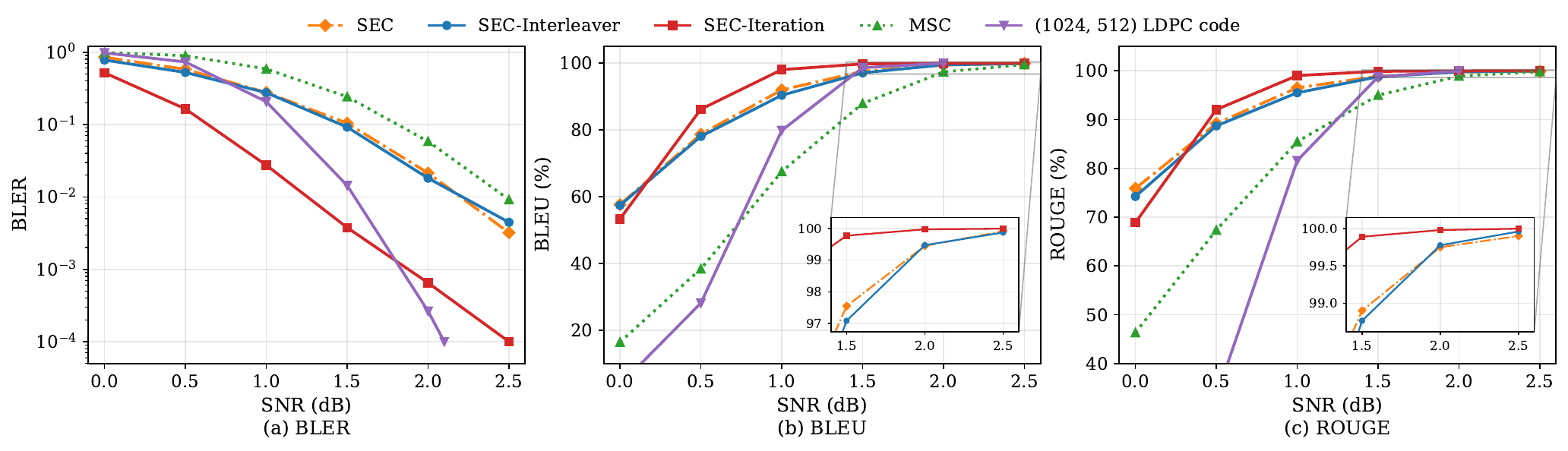}
    \vspace{-0.8em}
   \caption{BLER, BLEU, and ROUGE comparison of SEC-Interleaver, SEC-Iteration, SEC~\cite{sec}, MSC and LDPC code over the AWGN channel.}
   \label{fig:main_result}
   \vspace{-0.8em}
\end{figure*}

\subsubsection{Benchmarks}
We evaluate five schemes, all MSC schemes using $\mathcal{C}_{\mathrm{MSC}}(128,64)$~\cite{sld} constructed from the 5G NR polar-code generator matrix in systematic form, with $q=8$ segments per 512-bit sentence:
\begin{itemize}
    \item \textbf{MSC} \cite{sec}: per-segment order-4 OSD decoding without semantic processing.
    \item \textbf{LDPC code}: the entire 512-bit sentence is encoded using a rate-1/2 5G NR LDPC code with 1024 transmitted bits and decoded by BP with up to 80 iterations.
    \item \textbf{SEC} \cite{sec}: the MSC pipeline followed by a single-pass BART reconstruction.
    \item \textbf{SEC-Interleaver:} MSC with character-level interleaving and a single SEC pass.
    \item \textbf{SEC-Iteration:} the proposed iterative SEC framework.
\end{itemize}

\begin{figure}[t]
   \centering
   \includegraphics[width=0.99\linewidth]{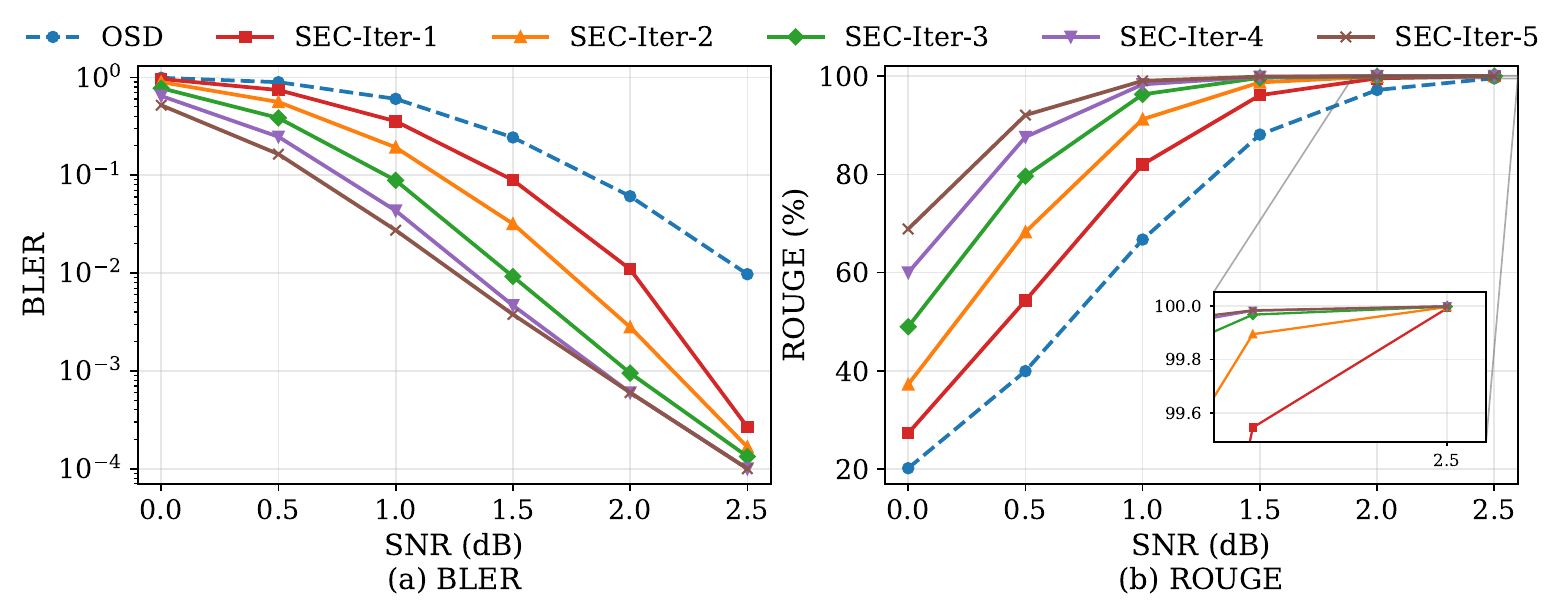}
   \vspace{-1.5em}
   \caption{BLER and ROUGE of SEC-Iteration across five iterations.}
   \label{fig:iteration_result}
   \vspace{-1em}
\end{figure}

\subsubsection{Evaluation Metrics}
We report sentence-level BLER, in which a block error occurs whenever the recovered sentence differs from the original after all processing. Semantic fidelity is measured by BLEU \cite{bleu} and ROUGE-L \cite{rouge}, evaluating $n$-gram precision and longest-common-subsequence similarity, respectively. ROUGE-L is denoted by ROUGE for brevity.


\subsection{Performance Comparison}
\label{subsec:performance}

Fig.~\ref{fig:main_result} compares the proposed SEC-Interleaver and SEC-Iteration schemes against the SEC~\cite{sec} and MSC baselines in terms of BLER, BLEU, and ROUGE. Fig.~\ref{fig:iteration_result} further examines the convergence behavior of SEC-Iteration across iterations.

SEC-Interleaver maintains a coding gain of approximately $0.4$~dB over MSC across the SNR range while substantially improving semantic fidelity, lifting the BLEU and ROUGE scores at $0$~dB from $16.4$ and $46.3$ to $57.4$ and $74.2$. Compared with single-pass SEC, the interleaver alone offers no significant benefit under single-pass decoding, because recovering a sentence requires correcting all erroneous characters, and dispersing them does not make this joint task easier for a single BART pass.

SEC-Iteration achieves the best BLER performance across the entire SNR range, outperforming SEC by approximately $1$~dB and MSC by approximately $1.5$~dB, with BLER reductions often exceeding an order of magnitude. At $0$~dB, SEC-Iteration reduces the BLER from $0.850$ (SEC) to $0.519$. It also outperforms the $(1024,512)$ LDPC baseline below approximately $1.75$~dB and reaches a BLER of $10^{-4}$ at $2.5$~dB, requires approximately 0.4 dB more SNR than LDPC at BLER, while SEC remains at $3.2\times10^{-3}$. In terms of semantic fidelity, SEC-Iteration is slightly inferior to SEC below $0.2$~dB because it returns the final OSD output rather than the subsequent SEC reconstruction. The OSD output is determined by the channel observations, whereas the SEC reconstruction is often more fluent but may overwrite correctly decoded characters with contextually plausible alternatives. It also substantially outperforms LDPC at low SNR, where LDPC achieves only $3.6\%$ BLEU and $8.3\%$ ROUGE at $0$~dB. Beyond $1$~dB, SEC-Iteration maintains both metrics above $99\%$, indicating highly reliable semantic recovery.

\vspace{-0.7em}

\subsection{Ablation Study}
\label{subsec:ablation}

We validate the two design choices of the character-level reliability verification module through targeted ablations at an SNR of $1.5$~dB, where MSC achieves a BLER of $0.247$.

\textbf{Rule~1:} Removing this rule, i.e., pinning every output character by its position regardless of the edit type, increases the BLER from $0.00314$ to $0.10514$. Pinning unchanged characters consumes the residual decoding capacity that would otherwise be used to correct genuine errors; moreover, any insertion or deletion misaligns all subsequent characters, so that the pinned values are largely incorrect and severely degrade the next iteration.

\textbf{Rule~2:} Removing the reliability threshold and admitting all substitution candidates raises the BLER from $0.00314$ to $0.00478$, an increase of approximately $52\%$. The empirical accuracy of substitution candidates drops from $99.4\%$ for $\rho_{r}\geq T_{\mathrm{h}}$ and $98.1\%$ for $T_{\mathrm{low}}\leq\rho_{r}<T_{\mathrm{h}}$ down to $83.9\%$ for $\rho_{r}<T_{\mathrm{low}}$, which reveals an inherent trade-off in threshold selection: lowering the threshold admits more correct candidates but also introduces more corrupted ones.

\begin{table}[t]
\centering
\caption{Per-sentence latency (ms) at different SNRs.}
\vspace{-0.8em}
\label{table:latency}
\renewcommand{\arraystretch}{1.15}
\begin{tabular}{@{}l | c c c c c c@{}}
\toprule
SNR (dB) & 0 & 0.5 & 1 & 1.5 & 2 & 2.5 \\
\midrule
MSC             & 8.3   & 8.3   & 8.3   & 8.3   & 8.3   & 8.3   \\
SEC         & 45.7  & 45.7  & 45.7  & 45.7  & \textbf{45.7}  & \textbf{45.7}  \\
SEC-Iteration        & \textbf{235.1} & \textbf{213.0} & \textbf{154.1} & \textbf{78.9}  & 34.4  & 14.0  \\
\bottomrule
\end{tabular}
\vspace{-1em}
\end{table}

\vspace{-0.7em}
\subsection{Latency Analysis}
We measure the software-level per-sentence latency using the PyTorch-based Sionna implementation, averaged over 550 test sentences.  For MSC, the parallel OSD decoding takes about $8.3$~ms, while one SEC pass adds $37.4$~ms, giving a total of $45.7$~ms per sentence for SEC. Table~\ref{table:latency} reports the per-sentence latency of the proposed iterative scheme across SNR. Unlike the fixed cost of MSC and SEC, the proposed scheme has an SNR-dependent latency: the iterative refinement terminates early once no unreliable segment remains, so higher SNR sentences require fewer iterations on average. At low SNR ($0$~dB), where most sentences trigger the full iteration budget, the average latency reaches $235.1$~ms. As SNR increases, an increasing fraction of sentences converges within fewer iterations, and the average latency drops sharply. At $2.5$~dB, the proposed scheme takes only $14.0$~ms per sentence, which is markedly lower than single-pass SEC and approaches the latency of MSC alone, while achieving substantially better BLER as shown in Fig.~\ref{fig:main_result}.


\section{Conclusion}
\label{sec:conclusion}
This paper proposed an iterative SEC framework in which language-model inference and channel decoding are alternately refined over successive rounds for short-block-code transmission of natural-language text. The proposed framework first incorporates a character-level interleaver to spread segment-level burst errors. Then, a verification module converts language-model reconstructions into semantic feedback by retaining only the characters consistent with the channel observations. The resulting feedback is incorporated into the subsequent decoding pass. Simulation results show approximately $1.5$~dB BLER gain over MSC and approximately $1$~dB gain over single-pass SEC.


\bibliographystyle{IEEEtran}
\bibliography{IEEEabrv, refs}

\end{document}